\documentclass[epj]{svjour}
\usepackage{graphics}
\begin{document}
\title{Unexpected transparency in the scattering of fragile 
$^{6}$Li and $^{6}$He Nuclei}
\author{F. Michel\inst{1} \and S. Ohkubo\inst{2}
}                     
%
%
\institute{Universit\'{e} de Mons-Hainaut, Place du Parc, 20,
B-7000 Mons, Belgium  
 \and Department of Applied Science and Environment,
 Kochi Women's University, Kochi 780-8515, Japan }
\date{Received:  10 August 2006 / Revised version: 20 August 2006}
%
\abstract{
  It is found that the scattering of the fragile nucleus $^6$Li from $^{12}$C 
and $^{16}$O  is     unexpectedly  transparent.
 It is shown  that the internal-wave 
contribution is significantly large in the  scattering,
 which suggests that some transparency could persist in the scattering 
involving the  fragile nucleus  $^6$He. 
\PACS{
      {25.60.Bx}{Elastic scattering}   \and
      {24.10.Ht }{Optical and diffraction models  }
      { 25.60.-t}{ Reactions induced by unstable nuclei}
     } 
} 
\maketitle


The nuclear rainbow and the Airy structure in elastic scattering 
are  observed when absorption is incomplete.
In the  scattering involving  magic nuclei like $\alpha$ particle, 
 $^{16}$O and  $^{40}$Ca  absorption is weak and nuclear rainbow has been
 typically
observed in the  $\alpha$+$^{16}$O,  $\alpha$+$^{40}$Ca  \cite{Michel1998} and
 $^{16}$O+ $^{16}$O systems \cite{Khoa2000}. For these systems nucleus-nucleus interaction
potential has been determined up to the internal region, which made it 
possible to study  the 
 cluster structure of the composite systems $^{20}$Ne, $^{44}$Ti
 \cite{Michel1998} and $^{32}$S \cite{Ohkubo2002}, 
respectively.   For a fragile
projectile  absorption becomes much stronger.  However we show that
the Airy
structure is  observed in the scattering involving a  fragile nucleus
 like $^6$Li 
and absorption is incomplete.
\begin{figure}[ht]
\resizebox{0.5\textwidth}{!}{%
  \includegraphics{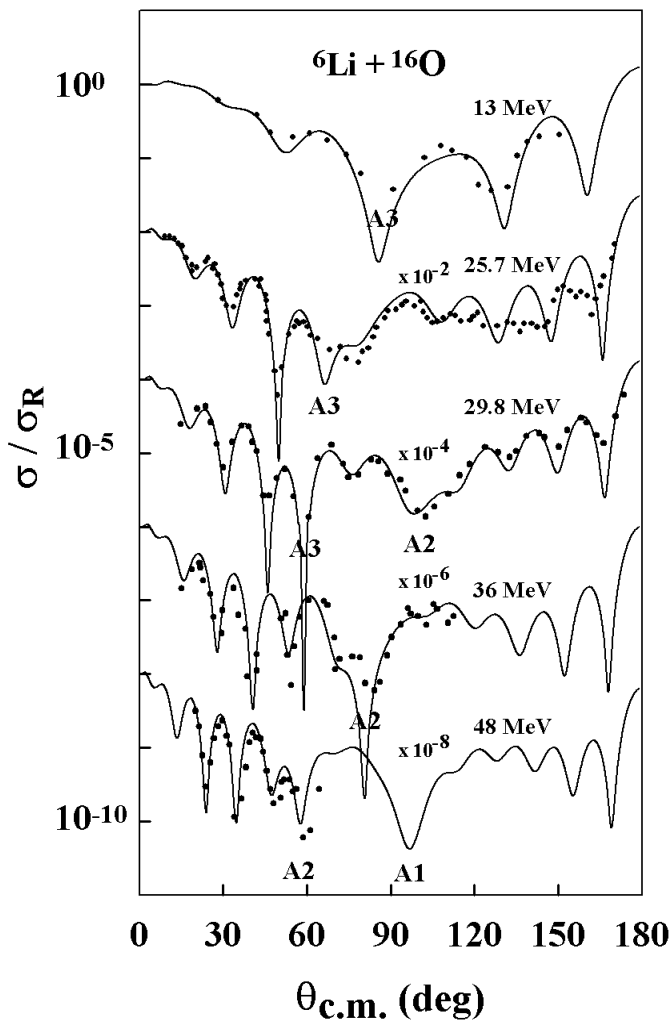}
}
\caption{Comparison of the  optical model $^6$Li+$^{16}$O cross
 sections (solid lines) and the 
experimental angular distributions (filled circles).
The potential parameters used are
$V_0$=316.2 MeV, $R_R$=3.319 fm, $a_R$=0.7 fm, $R_I$=5.822 fm, 
and $a_I$=0.613 fm in the conventional notation. $W_0$ is adjusted to 
7, 10, 11.5, 13.5 and 16 MeV for $E_L$=13, 25.7, 29.8, 36 and 48 MeV,
 respectively. Airy minima are also indicated. 
}
\label{fig:1}       
\end{figure}


\begin{figure}
\resizebox{0.5\textwidth}{!}{%
  \includegraphics{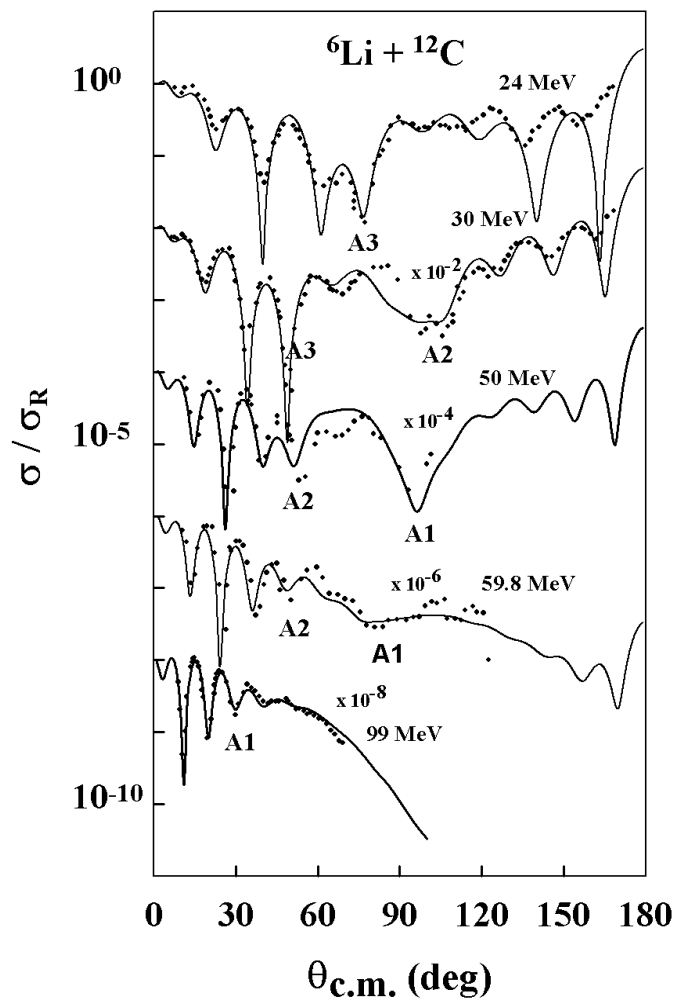}
}
\caption{Comparison of the global optical model $^6$Li+$^{12}$C cross
 sections (solid lines) and the 
experimental angular distributions (filled circles).
 Airy minima are also indicated.}
\label{fig:2}       
\end{figure}
\indent $^6$Li~+~$^{16}$O  elastic scattering shows Anomalous Large Angle Scattering
 (ALAS) \cite{Michel1998} as shown in fig.~1. 
The experimental data are well reproduced by an optical potential
 with a 
 squared Woods-Saxon form factor for real and imaginary potentials.
This potential is deep 
  similar to a folding model potential and the  global optical 
potential for the $\alpha$+$^{16}$O system  \cite{Michel1998}.
 By decomposing the  calculated scattering amplitude at $E_L$=25.7 and
 29.8 MeV 
into the internal-wave, which penetrates deep into the 
internal region of the potential,  and the  barrier-wave 
reflected at the barrier \cite{Michel2000}, it is clearly understood that
  the large cross sections at backward 
angles are due to the internal-wave contributions. This shows that
this scattering is transparent similar to the $\alpha$+$^{16}$O system
although $^6$Li  is  weakly bound.
By decomposing the calculated scattering amplitude into farside and nearside
components, the angular position of the Airy minimum is identified
 as shown in fig.~1.
For example, the broad dip at  $\theta_{c.m.}\simeq$ 100$^\circ$
 in the angular distribution for  $E_L$=29.8 MeV 
 is an
 Airy minimum of second order 
 $A2$.  For this system unfortunately there are no systematic experimental
 angular distributions at  high energy region where a typical fall-off of
 the cross sections appears.

 For the $^6$Li~+~$^{12}$C system there are systematic angular distributions 
 for incident energies ranging from a few  MeV to 318~MeV. 
 We have analyzed   $^6$Li~+~$^{12}$C elastic
 scattering  in  the optical model. In  fig.~2 the  experimental data 
are well reproduced by a global optical potential. 
The  Airy minima
  are clearly identified in the angular distributions.
Although the target nucleus $^{12}$C  is deformed and easily excited
 than the $^{16}$O nucleus,  
 this  fragile projectile  displays a 
surprising transparency in the scattering in spite of
 the important breakup  effects.  
This ALAS may be related to the $\alpha$-cluster structure  of the projectile nucleus
 and/or the formation of  resonances in the intermediate system.  
 We note that the energy evolution of the Airy minimum 
for the $^6$Li~+~$^{16}$O  and $^6$Li~+~$^{12}$C systems shows a 
similar behaviour, which is also similar  to  $\alpha$+$^{16}$O scattering 
 \cite{Michel1998,Michel2005}.
 
The transparency of the $^6$Li~+~$^{12}$C system can  be further 
confirmed by studying
the inelastic scattering.
In the coupled channel calculation 
of $^{12}$C($^6$Li,$^6$Li')$^{12}$C$^*$(J$^\pi = 2^+$,
 E$_x = 4.44$ MeV) inelastic scattering by using a collective form factor
for  $^{12}$C  the experimental  angular distributions  at $E_L$= 24 and 30 MeV
are well reproduced.
By decomposing the inelastic scattering amplitude into internal-wave and 
barrier-wave
 components in the frame of the DWBA \cite{Michel2005},
 it is found that the contribution 
of the internal-wave amplitude
is very large and the rise of the cross sections  beyond $\theta_{c.m.}$=80$^\circ$
comes  entirely from the internal-wave contributions.

  We have also investigated the angular distribution
 of elastic 
  $^6$He~+~$^{12}$C scattering data  at $E_L$=18 MeV \cite{Milin2004}.
Our potential obtained in the analysis of 
$^6$Li~+~$^{12}$C  scattering at the corresponding energy
 can reproduce the observed angular distribution,
which extends up to $\theta_{c.m.}\simeq$ 85$^\circ$.
The barrier-wave and internal-wave decomposition of the scattering
amplitude shows that significantly large contribution to the backward 
angles comes from the internal waves, which
 suggests that transparency could persist in this system.
From the similarity between the $^6$Li+$^{12}$C system and the 
$^6$Li~+~$^{16}$O system it is naturally expected  that the
 scattering for the $^6$He+$^{12}$C  and $^6$He~+~$^{16}$O systems 
would show a similar behaviour. Therefore 
it would highly desired to  measure the angular distribution
for $^6$He~+~$^{16}$O  scattering  as well as $^6$He+$^{12}$C scattering
 up to large angles in the energy range in fig.~2, which would ascertain 
whether  transparency persists in the scattering of  $^6$He from $^{12}$C
and  $^{16}$O.   In $\alpha$ particle scattering from $^{16}$O, $^{15}$N and 
$^{14}$C,  the angular distributions show almost similar behaviour under weak
 absorption.
In fact, the $\alpha$-nucleus potential for these nuclei is very 
similar each other and can be reproduced 
 by a double folding model. 
 A similar situation may be expected for $^6$He scattering from $^{16}$O,
 $^{15}$N and $^{14}$C, which will make it possible to determine the
 $^6$He-nucleus potential for these systems.

One of the authors (S.O.) has been supported by a Grant-in-Aid for Scientific 
Research of the 
Japan Society for Promotion of Science (No. 16540265) and the Yukawa Institute
for Theoretical Physics.

%
%

%

%

\begin{thebibliography}{}
%
%
\bibitem{Michel1998}
 F. Michel, S. Ohkubo, and 
 G. Reidemeister, 
 Prog. Theor. Phys.  {\textbf 132}, (1998) 7. 
\bibitem {Khoa2000} 
Dao T. Khoa, {\it et al.} 
 {\textbf A672},  (2000) 387.
\bibitem {Ohkubo2002} 
S. Ohkubo and K. Yamashita, 
Phys. Rev. {\textbf C 66},  (2002) 021301.
\bibitem{Michel2000}
 F. Michel, F. Brau, G. Reidemeister, 
and S. Ohkubo, Phys. Rev. Lett. {\textbf 85}, (2000) 1823.
\bibitem{Michel2005}
 F. Michel and S. Ohkubo,
Phys. Rev. {\textbf C 72},  (2005) 054601.
\bibitem{Milin2004}
 M. Milin {\it et al.}, 
  Nucl. Phys. {\textbf A730}, (2004) 285.

\end{thebibliography}
%

\end{document}